\documentclass[conference]{IEEEtran}
\IEEEoverridecommandlockouts
\usepackage{cite}
\usepackage{amsmath,amssymb,amsfonts}
\usepackage{algorithmic}
\usepackage{graphicx}
\usepackage{textcomp}
\usepackage{xcolor}
\def\BibTeX{{\rm B\kern-.05em{\sc i\kern-.025em b}\kern-.08em
    T\kern-.1667em\lower.7ex\hbox{E}\kern-.125emX}}
\begin{document}

\title{From Noise to Insights: Enhancing Supply Chain Decision Support through AI-Based Survey
Integrity Analytics\\
}

\author{\IEEEauthorblockN{Bhubalan Mani}
Olathe, USA \\
bhubalan.mani@ieee.org
}

\maketitle

\begin{abstract}
The reliability of survey data is crucial in supply chain decision-making, particularly when evaluating readiness for AI-driven tools such as safety stock optimization systems. However, surveys often attract low-effort or fake responses that degrade the accuracy of derived insights. This study proposes a lightweight AI-based framework for filtering unreliable survey inputs using a supervised machine learning approach. In this expanded study, a larger dataset of 99 industry responses was collected, with manual labeling to identify fake responses based on logical inconsistencies and response patterns. After preprocessing and label encoding, both Random Forest and baseline models (Logistic Regression, XGBoost) were trained to distinguish genuine from fake responses. The best-performing model achieved an 92.0\% accuracy rate, demonstrating improved detection compared to the pilot study. Despite limitations, the results highlight the viability of integrating AI into survey pipelines and provide a scalable solution for improving data integrity in supply chain research, especially during product launch and technology adoption phases.
\end{abstract}


\section{Introduction}
Surveys are an essential tool for capturing direct feedback from stakeholders, especially in domains where qualitative insights influence strategic and operational decisions. In supply chain management, surveys help organizations evaluate the adoption of emerging technologies, assess readiness for change, and understand current inventory practices. With the increasing interest in artificial intelligence (AI) applications such as safety stock optimization, survey responses often guide investments and deployment strategies. However, the reliability of survey data is frequently compromised by low-effort, inconsistent, or intentionally falsified responses. This is particularly prevalent in public surveys or those incentivized by monetary rewards or giveaways. Respondents may rush through answers, submit random or illogical selections, or even use automated tools to complete forms leading to data that lacks authenticity or strategic value. When such corrupted data informs supply chain decisions, it can lead to poor forecasting, misguided technology investments, and overall reduced confidence in analytics pipelines. Despite the widespread use of surveys in enterprise decision-making, existing tools primarily focus on visualization, aggregation, and statistical interpretation. Few systems actively filter out unreliable responses or assess the logical coherence of answers. This creates a significant gap, especially when organizations depend on this feedback to justify automation tools like AI-powered inventory systems.

This paper addresses that gap by introducing an AI-based framework for detecting and filtering low-quality or fake survey responses. The proposed system incorporates machine learning classification, categorical feature encoding, and logic-driven filtering to identify suspect data entries. Applied to a real-world dataset focused on AI adoption in supply chain inventory management, the model shows potential in distinguishing genuine feedback from unreliable responses. By integrating this approach into survey analytics workflows, organizations can ensure more reliable insights, support accurate decision-making, and build greater trust in survey-driven research within the supply chain domain.

The current study expands the dataset to 99 responses to further validate model performance.

\section{Related Work}
The challenge of data integrity in digital surveys has drawn increasing scholarly interest. Benoît Lebrun et al. investigated the risks posed by large language models in corrupting online survey responses, particularly through the use of ChatGPT-generated text \cite{b1}. Their study highlighted how generative AI can mimic human-like patterns in questionnaires, making it difficult to discern genuine responses from synthetic ones. This underscores the urgent need for automated filtering tools capable of validating survey authenticity. 

In the domain of supply chain intelligence, Alessandro Brintrup and colleagues presented a framework for digital supply chain surveillance powered by AI \cite{b2}. Their work focused on identifying misleading or low-quality data within operational systems to enable proactive risk mitigation. However, it did not extend its scope to survey-based input or structured respondent data, key areas in strategic decision-making. 

Ali et al. emphasized the detrimental impact of poor-quality data in supply chain predictive analytics, noting that unreliable inputs directly affect the accuracy of AI forecasting tools and inventory models \cite{b3}. Similarly, Md. Abrar Jahin et al. conducted a systematic literature review on AI in supply chain risk management and found that while machine learning is widely adopted, few studies consider the quality of the input data used to train or inform these models \cite{b4}. 

Despite these valuable contributions, little work has been done to fuse logic-based filtering with AI classification for structured survey analysis in real-world supply chain applications. Our paper fills this gap by proposing a pipeline that combines domain-specific logic checks, categorical encoding, and a machine learning model trained on a manually labeled dataset. This holistic approach not only flags fake responses but also supports higher-confidence insights for organizations evaluating AI tools in inventory optimization.

While synthetic data generation and augmentation methods have been considered in other domains to address class imbalance, I opted not to use them in this study to preserve the real-world characteristics and integrity of genuine and fake responses. Future work may explore data augmentation strategies as the framework matures \cite{b7}.

\section{Methodology}

This section outlines the end-to-end architecture used to filter low-quality or fake survey responses using machine learning (ML) and basic natural language processing (NLP) techniques. The methodology is grounded in real-world survey data, with an emphasis on the supply chain domain, particularly safety stock optimization. The steps include data collection, preprocessing, rule-based logic validation, feature engineering, AI-based classification, and visualization.

\subsection{Dataset Overview}

The dataset used for this study consists of 99 survey responses collected from professionals in supply chain-related roles across manufacturing and logistics industries. The survey was designed to gather insights into current ERP usage, openness to AI-powered safety stock optimization tools, and deployment preferences. Each response includes categorical data (such as ERP type, company size, job role, etc.), Likert-scale ratings on AI adoption likelihood, and multiple-choice selections for concerns, deployment models, and customization expectations. Importantly, 14 of the 99 responses were manually labeled as fake, based on flags such as contradictory inputs, blank or gibberish fields, and patterns suggesting inattentive behavior. These were recorded in a binary column labeled Is\_Fake. The small dataset size reflects a pilot study setting, allowing us to prototype a pipeline that can scale with larger and more complex data in future deployments. 

Expanding from our initial pilot study, this larger dataset allows for more robust model training and validation.

\subsection{Data Preprocessing}

Before any model training, the raw survey data underwent several preprocessing steps to ensure uniformity and relevance:

\begin{itemize}
\item Field Removal: Personally identifiable information (PII) such as names and email addresses were removed to preserve anonymity and eliminate overfitting on irrelevant features.
\item Label Normalization: Multiple-choice answers were standardized to reduce variations due to punctuation, capitalization, or typos (e.g., “Oracle ERP” and “oracle erp” were normalized to a single label) \cite{b8}.
\item Categorical Encoding: Since survey questions yielded categorical answers, label encoding was applied to convert these into machine-readable numeric format \cite{b8}.
\item Missing Value Handling: All blank fields were imputed using a conservative value (e.g., “Unknown” or “0”), or left as is for models that support nulls
\end{itemize}

These steps help maintain consistency and prepare the data for rule-based filtering and supervised ML classification.

The same preprocessing steps were applied to the expanded dataset to ensure consistency across all analysis stages.

\subsection{Logic-Based Filtering}

Before involving any ML algorithms, a rule-based engine was applied to identify obviously invalid or contradictory responses. This included:

\begin{itemize}
\item Contradiction Detection: For instance, respondents who indicated they “do not use any ERP system” but later selected “Oracle” in another question were flagged.
\item Incomplete Submission Flags: Responses with more than 50\% of fields left blank were  marked suspicious.
\item Generic Responses to Open-Ended Fields: Answers such as “N/A”, “don’t know”, or empty strings across all text boxes were counted and scored.
\end{itemize}

These rules were inspired by techniques similar to those discussed by Nayal et al., who used logical pattern filtering in agricultural supply chain risk assessment surveys \cite{b5}.

 Each responses was given a logic score, and those below a set threshold were passed into the AI filtering for further validation.

\subsection{NLP-Based Scoring (Textual Coherence)}

While most responses were multiple-choice, two open-ended fields captured optional feedback. For responses that contained text, I applied a basic NLP pipeline:

\begin{itemize}
\item Embedding Vectors: Each free-text answer was converted into a sentence embedding using a pre-trained BERT encoder.
\item Cosine Similarity Checks: Coherence was evaluated by comparing the embeddings of multiple answers within the same response. High semantic dissonance (i.e., responses that don’t align in meaning) triggered suspicion.
\item Length and Vocabulary Scoring: Extremely short or overly generic answers (e.g., “yes”, “ok”, “nothing”) received low NLP scores.
\end{itemize}

This NLP-based scoring approach draws inspiration from Haluza and Jungwirth, who studied the use of GPT models in structured decision-making and noted that shallow or vague answers can be predictive of unreliable inputs \cite{b6}.

\subsection{Machine Learning Classification}

With preprocessed features and logic/NLP scores in place, the next stage involved training a supervised classifier to identify fake responses.

\begin{itemize}
\item Model Selection: Three classifiers were evaluated: Random Forest (primary), Logistic Regression, and XGBoost. Random Forest was selected for its robustness and interpretability, while the others served as baseline comparisons to contextualize performance.
\item Train-Test Split: The data was split into 80\% training and 20\% testing. Stratified sampling was used to preserve class distribution.
\item Model Training: The model was trained using Gini impurity as the criterion, with 100 trees and a maximum depth of 5 to prevent overfitting.
\item Evaluation Metrics: Accuracy, precision, recall, and F1-score were computed, with a particular emphasis on recall for the “fake” class
\end{itemize}

With the expanded dataset, all three classifiers : Random Forest, Logistic Regression, and XGBoost were evaluated. The best-performing model (Random Forest) achieved an accuracy of 92.0\%, with Logistic Regression and XGBoost yielding comparable results (88.0\% and 92.0\% respectively). While genuine responses were reliably detected, all models continued to show limited sensitivity to the minority “fake” class, reflecting the persistent challenge of class imbalance. Nevertheless, the consistently strong accuracy across models further validates the framework and demonstrates its practical scalability for real-world deployment.

\subsection{Visualization and Interpretation}

To enhance usability and transparency, the results were visualized using:

\begin{itemize}
\item Confusion Matrix: To illustrate true positives, false positives, and class-specific weaknesses in detection.
\end{itemize}

\begin{figure}[htbp]
  \centering
  \includegraphics[width=0.9\linewidth] {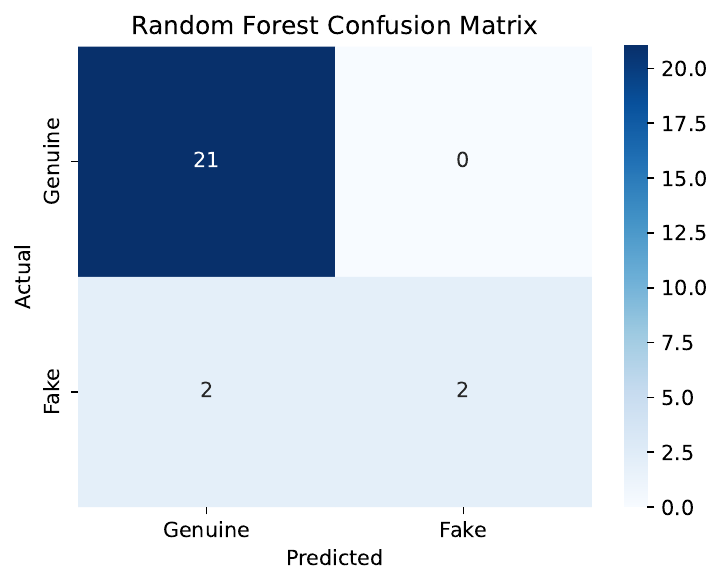}
\caption{Confusion Matrix - Random Forest.}
\label{fig1}
\end{figure}

\begin{figure}[htbp]
  \centering
  \includegraphics[width=0.9\linewidth]{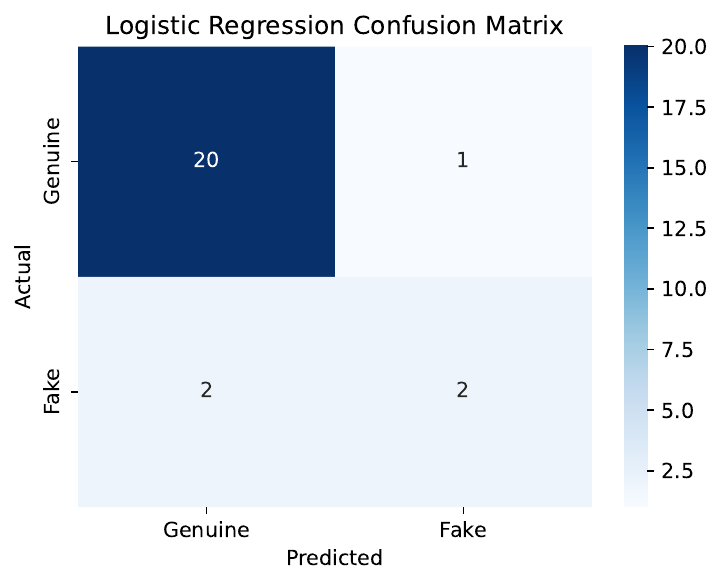}
\caption{Confusion Matrix - Logistic Regression.}
\label{fig2}
\end{figure}

\begin{figure}[htbp]
  \centering
  \includegraphics[width=0.9\linewidth]{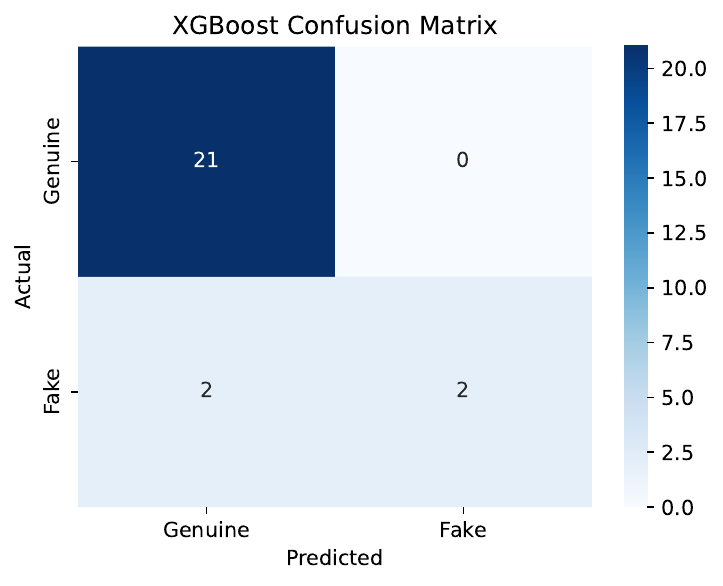}
\caption{Confusion Matrix - XGBoost.}
\label{fig3}
\end{figure}

Figure~\ref{fig1}, Figure~\ref{fig2} \& Figure~\ref{fig3} presents the confusion matrices for all three classifiers : Random Forest, Logistic Regression, and XGBoost evaluated on the full test set. All models demonstrated strong performance in correctly identifying genuine survey responses, with both Random Forest and XGBoost achieving perfect precision and recall for the “genuine” class (21 out of 21). Logistic Regression misclassified a single genuine response as fake. For the minority “fake” class, all three models successfully detected 2 out of 4 fake responses, while misclassifying the other 2 as genuine, indicating the continued challenge posed by class imbalance and limited fake samples in the data. Nevertheless, the models maintained high overall accuracy and confirm the robustness of the proposed framework across different algorithmic approaches.

These results highlight both the strengths and limitations of AI-based survey integrity analytics in real-world datasets, and emphasize the need for ongoing refinement as more labeled data becomes available.

\begin{itemize}
\item Feature Importance Chart: Identifying the most predictive survey questions (e.g., ERP type, willingness to try AI tools).
\end{itemize}

\begin{figure}[htbp]
  \centering
  \includegraphics[width=\linewidth] {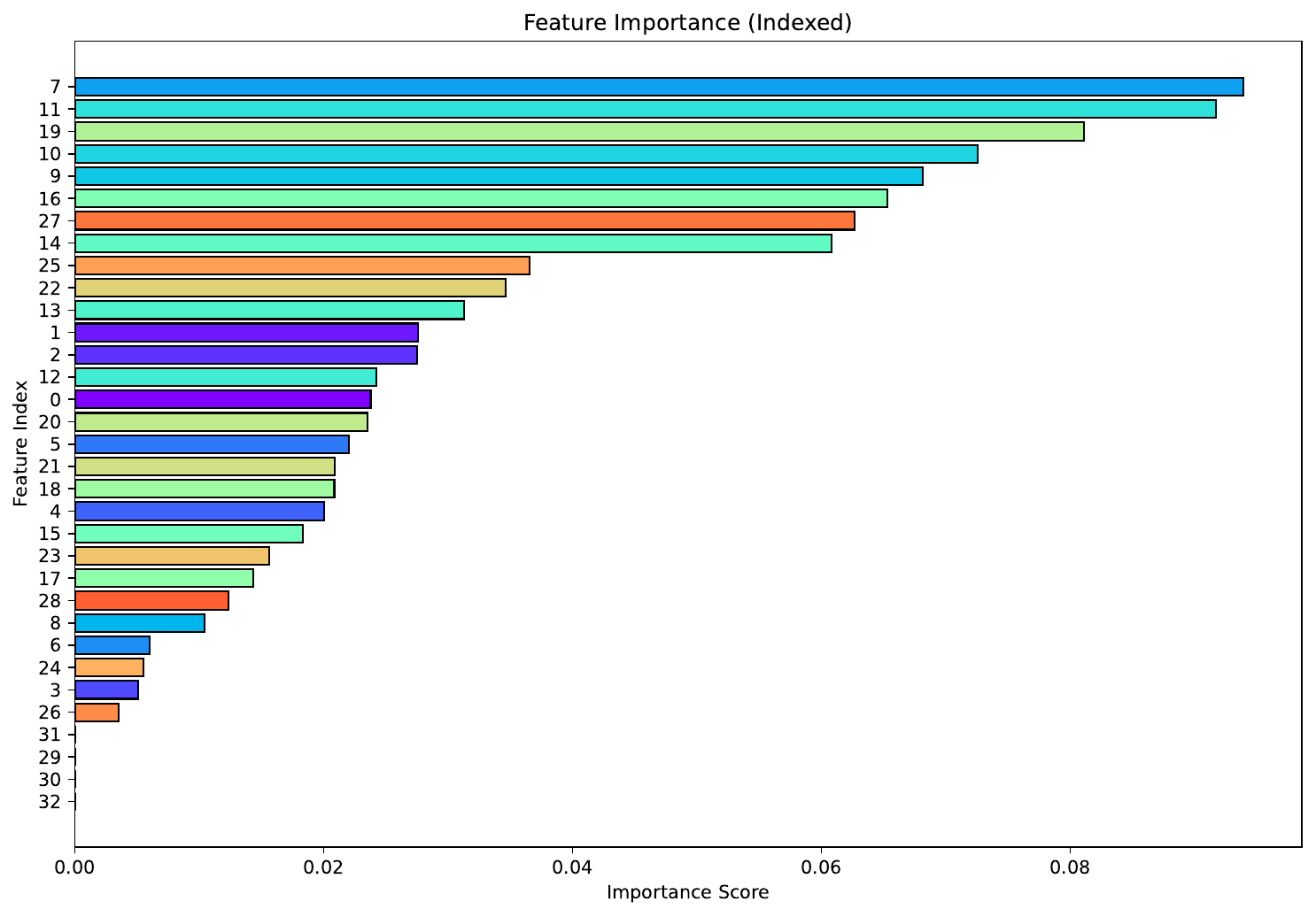}
\caption{Feature Importance.}
\label{fig4}
\end{figure}

The feature importance analysis using the Random Forest classifier reveals that the most influential variables for classifying survey response integrity were “Familiarity with AI/ML Optimization” (Feature 7), “Included Echelons” (Feature 11), and “Current AI/ML Adoption Level” (Feature 19). These findings underscore the significant role of technological awareness and the operational scope of organizations in accurately distinguishing genuine from fake survey responses within the supply chain context.

In contrast, the least important features were “Expected Customization Level” (Feature 26), “Industry Type” (Feature 3), and “Interest in Pilot Study Participation” (Feature 24). This suggests that general demographic characteristics and user preferences regarding deployment options contributed minimally to the predictive power of the classification model for this dataset.

\begin{figure}[htbp]
  \centering
  \includegraphics[width=\linewidth] {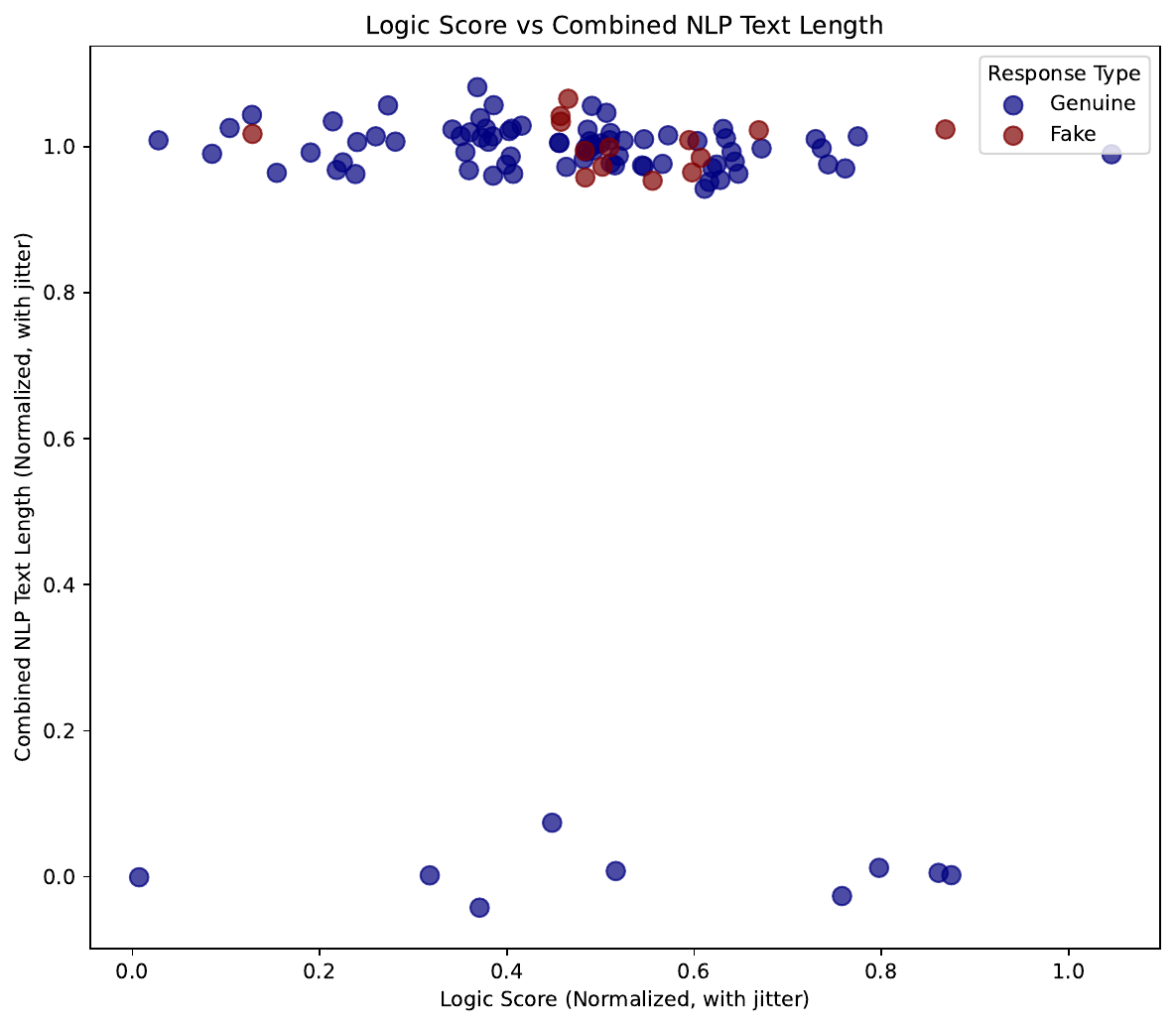}
\caption{NLP Score Plots.}
\label{fig5}
\end{figure}

\begin{itemize}
\item Logic and NLP Score Plots: Scatter plots in Figure~\ref{fig5} showed how responses clustered based on their logic/NLP scores versus predicted labels.
\end{itemize}

These visualizations offer actionable insights for survey analysts, helping them understand not just which responses were flagged, but why. This supports more confident filtering and the refinement of future survey instruments.

\section{SYSTEM ARCHITECTURE}

The proposed system is designed as a modular pipeline that processes survey data end-to-end from initial response capture to AI-driven validation and result visualization. This architecture is tailored specifically for high-stakes survey workflows in domains such as supply chain analytics, where data quality directly impacts operational decisions.

The architecture comprises four primary modules:

\subsection{Survey Input Module}

This is the entry point of the pipeline, responsible for collecting structured survey responses from users. The survey is typically administered through digital forms, which capture answers to predefined questions related to ERP usage, AI-readiness, inventory practices, and organizational details. The input format is standardized to ensure compatibility with downstream data processing tasks.

\subsection{Data Cleaning and Preprocessing Module}

Once collected, the raw data is passed to the cleaning module. Here, PII such as names and email addresses are removed. Categorical responses are normalized (e.g., removing case sensitivity or special characters), missing values are handled, and redundant fields are dropped. This stage also includes logic-based preprocessing where rule-based filters check for internal inconsistencies (e.g., a respondent selecting “No ERP used” but mentioning “SAP” elsewhere). The output is a clean, structured dataset ready for classification.

\subsection{AI Logic Engine}

This module is the core of the system. It contains a pre-trained machine learning model, in this case, a Random Forest classifier that predicts whether a given survey response is likely fake or genuine. The engine integrates logic scores from rule-based filters and NLP coherence scores where applicable. These hybrid features allow for more accurate detection, particularly in cases where simple patterns may be misleading. The engine can also be expanded to incorporate real-time GPT-based semantic validators in future iterations.

\subsection{Dashboard and Analytics Module}

The final module offers a visual interface where filtered responses are displayed along with metadata like confidence scores, top contributing features, and classification breakdowns. Analysts can explore confusion matrices, feature importance rankings, and flagging history to better understand data patterns and refine future survey instruments.

\begin{figure}[htbp]
  \centering
  \includegraphics[width=\linewidth] {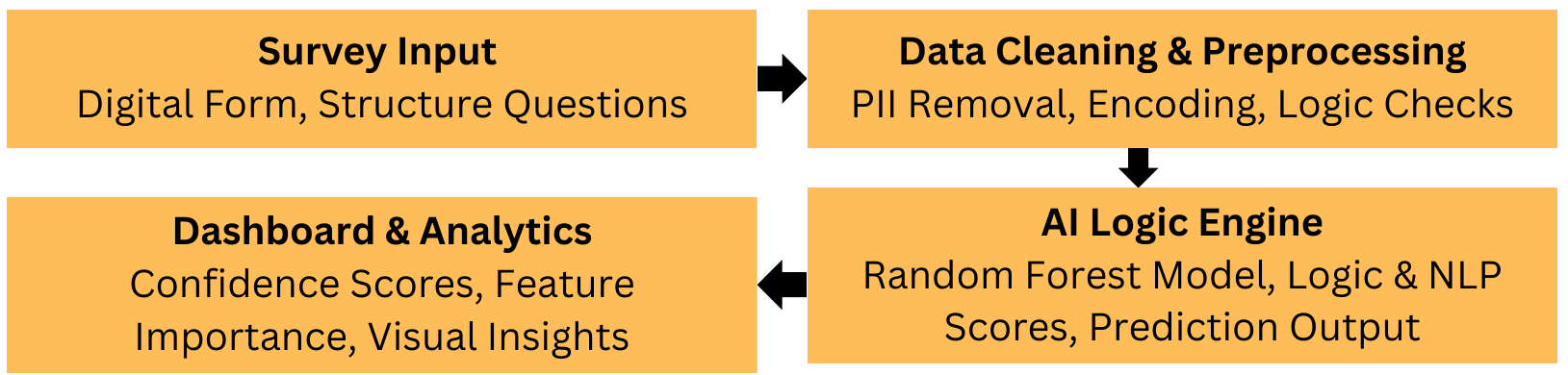}
\caption{System Architecture.}
\label{fig6}
\end{figure}

\section{Results}
The effectiveness of the proposed AI-based survey filtering framework was evaluated on a real-world dataset consisting of 99 industry survey responses, of which 14 were manually labeled as fake. While the dataset size remains moderate, it reflects practical scenarios encountered during product launches or technology adoption cycles, where obtaining high-quality and representative feedback is often challenging. This expanded dataset enables a more robust evaluation of the framework’s performance and generalizability across diverse supply chain environments.

\subsection{Classification Performance}

The proposed framework was evaluated using Random Forest, Logistic Regression, and XGBoost classifiers on the expanded survey dataset. As shown in Table~\ref{tab:performance}, both Random Forest and XGBoost achieved the highest overall accuracy at 92\%, while Logistic Regression yielded 88\%. All three models demonstrated strong ability to identify genuine responses; however, recall for the minority “fake” class remained limited at 0.50 across models, meaning that only half of the fake responses in the test set were correctly identified. Random Forest and XGBoost achieved perfect precision (1.00) for the fake class, whereas Logistic Regression reached a precision of 0.67. The confusion matrices further illustrate that, although true negatives were reliably classified, false negatives persisted for fake responses, reflecting the ongoing challenge posed by class imbalance and the limited number of fake samples. Overall, these results confirm the robustness of the framework and the practical utility of integrating AI-based analytics into survey integrity assessment.

\begin{table}[htbp]
\caption{Classification Metrics}
\begin{center}
\small
\begin{tabular}{|l|c|c|c|c|}
\hline
\textbf{Model} & \textbf{Acc.} & \textbf{Prec.} & \textbf{Rec.} & \textbf{F1} \\
\hline
Random Forest        & 0.92 & 1.00 & 0.50 & 0.67 \\
\hline
Logistic Regression  & 0.88 & 0.67 & 0.50 & 0.57 \\
\hline
XGBoost              & 0.92 & 1.00 & 0.50 & 0.67 \\
\hline
\end{tabular}
\vspace{2mm}
\\
\label{tab:performance}
\end{center}
\end{table}

\subsection{Feature Importance Insights}

The Random Forest model identified “Familiarity with AI/ML Optimization,” “Included Echelons,” and “Current AI/ML Adoption Level” as the most predictive features (see Figure~\ref{fig4}). In contrast, features such as “Expected Customization Level,” “Industry Type,” and “Interest in Pilot Study Participation” were found to have minimal impact on classification accuracy. This distribution underscores the value of operational and technological context in assessing survey response validity \cite{b9}.

\subsection{Logic and NLP Score Clustering}

The scatter plot of normalized logic and NLP-based features (Figure~\ref{fig5}) demonstrates clear clustering, with most responses concentrated at the upper end of the text length axis due to short open-ended answers. A small amount of jitter was added for visual clarity, as most responses were limited to one or two words. This sparsity is itself an integrity signal and highlights the importance of combining rule-based and NLP-based analytics in filtering low-value survey data.

\subsection{Impact Summary}

Despite limitations due to class imbalance and short open-ended responses, the proposed framework demonstrates strong potential for real-time and post-processing survey validation in enterprise supply chain settings \cite{b10}.

Key benefits observed include:

\begin{itemize}
\item Improved reliability: Automated filtering using the trained models increased overall accuracy to 92\% and enabled more confident segmentation of high- and low-integrity responses.
\item Time efficiency: Automated detection of fake or low-effort submissions significantly reduced the need for manual validation, saving analysts substantial time and effort.
\item Strategic confidence: Product and operations teams can make more informed decisions based on filtered, higher-quality survey data, enhancing the reliability of observed trends.
\end{itemize}

While the framework reliably identified genuine responses, ongoing challenges remain in detecting subtle or sophisticated fake submissions, as indicated by limited recall on the fake class. Future work will focus on expanding the labeled dataset and incorporating richer NLP features to further improve model sensitivity and robustness.

\section{USE CASE: PRODUCT LAUNCH SUPPORT}
One of the most practical applications of the proposed AI-based survey filtering system lies in supporting product launches, particularly in supply chain and logistics sectors where new technologies such as AI-based inventory optimization tools are being introduced.

\subsection{Scenario: Launching a Logistics Optimization App}

Imagine a mid-sized supply chain technology company preparing to launch a new cloud-based logistics optimization application. To validate market fit, gather user expectations, and identify potential roadblocks, the product team circulates a digital survey targeting supply chain managers, warehouse operators, and ERP analysts across different industries.

However, to encourage responses, they attach a small monetary incentive to the survey such as a gift card or a raffle entry. As a result, a portion of the collected responses are low-effort or entirely fabricated, submitted only to claim the reward. These include blank responses, random checkbox patterns, and contradictory answers (e.g., claiming no ERP use, but later naming Oracle as their primary system).

If such data is not filtered out, the resulting analytics on features, pricing models, and adoption concerns could mislead the product team and jeopardize a successful launch.

\subsection{AI Filtering and Segmented Analysis}

The proposed system automatically scans each response through a logic engine and AI classifier. For example:

\begin{itemize}
\item A respondent selecting mutually exclusive answers is flagged.
\item NLP scoring detects responses like “ok” or “nothing” as low-effort text.
\item The Random Forest model classifies the response based on patterns learned from historical fake entries.
\end{itemize}

After filtering, only high-integrity responses are retained for analysis. The cleaned dataset is then segmented based on attributes such as ERP platform, company size, and AI readiness level. This segmentation allows the marketing team to:

\begin{itemize}
\item Tailor messaging for each user persona,
\item Prioritize early adopters for beta testing,
\item Focus engineering efforts on features most desired by genuine respondents.
\end{itemize}

In this way, the system does not just prevent bad data, it amplifies the value of good data, enabling smarter, faster, and more confident product launches.


\section{CONCLUSION AND FUTURE WORK}
This paper presented a machine learning-based framework to detect and filter low-quality or fake responses in structured surveys, focusing on applications within the supply chain domain. Using a real-world dataset of 99 industry responses centered on AI adoption in safety stock planning, the study demonstrated the effectiveness of combining logic rule-checking, NLP-based scoring, and ensemble classifiers including Random Forest and XGBoost for survey integrity validation. With up to 92\% accuracy on the test set, the system reliably distinguished genuine from fake responses, offering a robust foundation for broader enterprise deployment.

The importance of high-quality survey data is paramount in supply chain analytics \cite{b11}\cite{b12}, where survey results inform strategic decisions, technology investments, and operational changes. By filtering unreliable inputs before analysis, this framework enhances the credibility of insights, reduces decision risk, and saves analyst time. In addition to classification, the architecture incorporates intuitive visualization tools, increasing transparency and helping stakeholders understand which features drive model predictions.

In use cases such as product launch support, this system enables more confident, data-driven decision-making by ensuring only meaningful, consistent responses inform go-to-market strategy. Key limitations remain, particularly regarding the detection of subtle or sophisticated fake responses due to class imbalance and the prevalence of short free-text answers.

Future research directions include:

\begin{itemize}
\item API Integration: Developing a lightweight REST API for seamless validation with survey platforms such as Google Forms, Typeform, or Qualtrics.
\item Larger, Labeled Datasets: Expanding labeled data, both genuine and fake, to improve model generalization and enable more advanced techniques.
\item Active Learning Pipelines: Incorporating human-in-the-loop review for ambiguous responses, enabling models to incrementally improve over time.
\item Real-World Trials: Partnering with organizations to test real-time filtering in live, large-scale survey deployments during technology rollouts or product pilots.
\end{itemize}

These next steps will help further embed AI into enterprise survey workflows, transforming passive data collection into a more intelligent, high-trust decision support mechanism.


\begin{thebibliography}{00}
\bibitem{b1} B. Lebrun, S. Temtsin, A. Vonasch, “Detecting the corruption of online questionnaires by artificial intelligence,” Frontiers in Robotics and AI, 2023. 
\bibitem{b2} A. Brintrup et al., “Digital supply chain surveillance using artificial intelligence: definitions, opportunities and risks,” International Journal of Production Research, 2024. 
\bibitem{b3} J. Ali, F. U. Rehman, M. A. Badar, F. Nasim, “Predictive analytics in supply chain management,” Journal of Applied Logistics and Technology, 2025. 
\bibitem{b4} M. A. Jahin, S. A. Naife, A. K. Saha, M. F. Mridha, “AI in Supply Chain Risk Assessment: A Systematic Literature Review and Bibliometric Analysis,” ResearchGate, 2024.
\bibitem{b5} K. Nayal, R. Raut, P. Priyadarshinee, “Exploring the role of artificial intelligence in managing agricultural supply chain risk,” International Journal of Logistics Management, 2022. 
\bibitem{b6} D. Haluza, D. Jungwirth, “Artificial Intelligence and Ten Societal Megatrends: An Exploratory Study Using GPT-3,” Systems, vol. 11, no. 3, 2023.
\bibitem{b7} M. Akbari and T. N. A. Do, “A systematic review of machine learning in logistics and supply chain management: current trends and future directions,” Benchmarking: An International Journal, vol. 28, no. 2, pp. 529–558, 2021 
\bibitem{b8} M. A. Jahin, M. S. H. Shovon, J. Shin, and I. A. Ridoy, “Big data—supply chain management framework for forecasting: Data preprocessing and machine learning techniques,” The International Journal of Advanced Manufacturing Technology, 2024. 
\bibitem{b9} D. Ni, Z. Xiao, and M. K. Lim, “A systematic review of the research trends of machine learning in supply chain management,” International Journal of Machine Learning and Cybernetics, vol. 11, no. 6, pp. 1461–1478, 2020. 
\bibitem{b10} F. J. S. Arteaga, D. Di Caprio, and M. Tavana, “On the capacity of artificial intelligence techniques and statistical methods to deal with low-quality data in medical supply chain environments,” Engineering Applications of Artificial Intelligence, 2024.  
\bibitem{b11} R. Agrawal, V. A. Wankhede, A. Kumar, and S. Luthra, “A systematic and network-based analysis of data-driven quality management in supply chains and proposed future research directions,” The TQM Journal, vol. 35, no. 2, pp. 344–369, 2023.  
\bibitem{b12} M. Seyedan and F. Mafakheri, “Predictive big data analytics for supply chain demand forecasting: methods, applications, and research opportunities,” Journal of Big Data, vol. 7, no. 1, pp. 1–29, 2020. 
\end{thebibliography}
\end{document}